\documentclass{optica-article}
\usepackage{subcaption}
\usepackage{gensymb}
\usepackage{booktabs}  
\usepackage{siunitx}   
\usepackage{adjustbox}

\journal{opticajournal} 

\articletype{Research Article}

\usepackage{lineno}

\begin{document}

\title{Photoionization in KTN deflectors by light in the near-infrared imaging window}

\author{Samuel Stanek,\authormark{1} Harishankar Jayakumar\authormark{2}, Christopher Warkentin\authormark{2}, James Leger\authormark{1}, and Aaron Kerlin\authormark{2,*}}

\address{\authormark{1}Department of Electrical and Computer Engineering, University of Minnesota, 200 Union Street SE, Minneapolis, MN 55455, USA\\
\authormark{2}Department of Neuroscience, Center for Magnetic Resonance Research, University of Minnesota, 2021 6th Street SE, Minneapolis, MN 55455, USA}

\email{\authormark{*}Corresponding author: akerlin@umn.edu} 


\begin{abstract*} 
Electro-optical deflectors (EODs) offer unparalleled scanning speed for laser-scanning microscopy and other applications, but suffer from limited deflection range. EODs based on potassium tantalate niobate (KTN) crystals feature some of the highest number of resolvable spots. These deflectors rely on internal electric fields generated by trapped electrons to enable beam scanning. However, visible light induces rapid photoionization of trapped charges, thus KTN-based deflectors are typically continuously recharged with a bias voltage that effectively limits the range of the deflector. Recent work has proposed the use of KTN-based EODs for biological imaging with infrared excitation light, but quantitative data on near-infrared photoionization is lacking. Here, we present quantitative measurements of photoionization in KTN deflectors across the NIR-I and NIR-IIa biological imaging windows (700--1300~nm), a range that is particularly important for deep tissue imaging and nonlinear microscopy. Using a two-beam polarization interferometer, we measured trapped charge density as a function of photon fluence. We observed that the photoionization rate decreases dramatically with increasing wavelength. The charge density decay curves exhibit multi-exponential behavior that cannot be explained by a single-trap model without recapture, indicating the presence of multiple trap species or substantial recapture. These measurements provide critical quantitative guidance for selecting operating wavelengths and charge-scan duty cycles for KTN-based EODs in near-infrared imaging applications.
\end{abstract*}

\section{Introduction}
\begin{figure}[htbp]
    \centering
    \includegraphics[width=\linewidth]{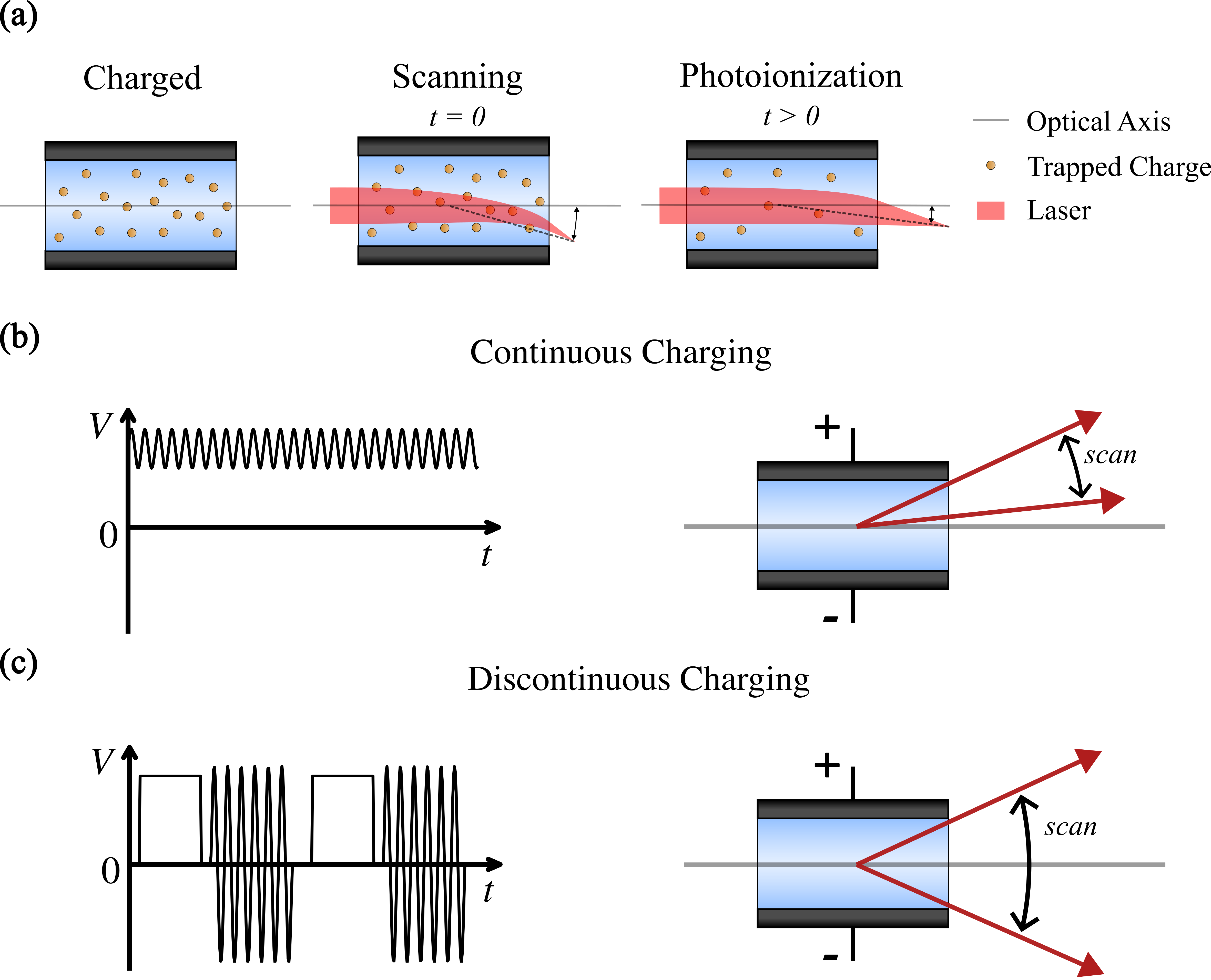}
    \caption{
    KTN deflector operation. (a) Prior to illumination, injected electrons are trapped inside the KTN. Initially, after charging, light can be scanned across a large angle. Photoionization reduces trapped charge density and scan angle over time. Continuous charging (b) uses a biased sinusoid, which replenishes charge but adds a bias angle, limiting the maximum geometric scan angle. Discontinuous charging (c) alternates between a DC pulse and a zero-bias sinusoid. This provides no bias angle and increases the maximum geometric scan angle.
    }
    \label{fig:intro}
\end{figure}

Electro-optical (EO) deflectors provide high-frequency, inertia-free scanning and have applications in imaging, telecommunications, and machining \cite{jimenez2023, yin2021, farinella2024}. As a scanning mechanism, the EO effect avoids the inertial speed limits inherent in galvanometer and acousto-optic deflectors. Potassium Tantalate Niobate (KTa$_{1-x}$Nb$_x$O$_3$; KTN) is an EO crystal with high manufacturability and mechanical robustness, exhibiting the highest known Kerr EO coefficient of any bulk material, which produces a refractive index change proportional to the square of the applied electric field \cite{wang2023, yagi2014}. For these reasons, KTN has attracted significant attention as a material of interest for high-speed laser scanning applications \cite{okabe2013, damodaran2018, sakamoto2014}. 

EO KTN-based deflectors rely on the electric field within the crystal for their operation. Most KTN deflectors use titanium electrodes to both apply voltage and inject free electrons from the cathode into the crystal \cite{nakamura2008, NTTAT_KTN_deflector, Toyoda2014}. Injected electrons move under the influence of the applied electric field, with some becoming trapped by defects in the crystal. The resulting space-charge distribution modifies the internal electric field, and through the Kerr effect, a quadratic refractive index profile is formed for light polarized parallel to the field. Applying an AC voltage to the charged crystal shifts the index profile along the electrode axis, deflecting the beam and enabling scanning \cite{miyazu2011}. During deflector operation, scanning beam photons can interact with trapped electrons, freeing them from traps. This process is referred to as photoionization. If a bias voltage is not continually applied to the deflector electrodes, photoionization can reduce the density of trapped charge over time. The reduction in trapped charge density will lead to a loss of scanning angle via reduced internal field. The charging, scanning, and photoionization processes are shown in Fig.~\ref{fig:intro}(a). 

Although applying a continuous bias voltage to the deflector electrodes can replenish photoionized charge, there are several disadvantages that make this approach unsuitable for some applications. When the device is continuously charged by a DC voltage, an angular offset is created that limits the scan angle through aperture clipping, as shown in Fig.~\ref{fig:intro}(b). Even in scanning applications that are not aperture limited, the bias voltage can limit the maximum allowable scanning voltage before piezoelectric strain damages the crystal.

Discontinuous charging, shown in Fig.~\ref{fig:intro}(c), avoids the aforementioned issues by alternating between a DC charging pulse and an unbiased AC sinusoid so that during scanning there is no DC voltage and thus no angular bias \cite{miyazu2011}. This permits higher scanning angles from both a geometric and electrical perspective. The lack of angular bias on the scanning beam is particularly in multi-pass configurations where an angular bias would accumulate and complicate deflector geometry \cite{jayakumar2025, yu2025}. Importantly, this operating mode requires that photoionization in the scanning period be minimized to preserve scan fidelity and reproducibility in imaging applications. 

Unfortunately, little is known about the photoionization behavior of trapped charge in KTN. It has been experimentally observed that UV light will deplete a KTN crystal of trapped charge within seconds \cite{yagi2014}. Blue 405~nm light has been used to intentionally excite electrons from traps \cite{zhu2018}. One application reported the use of discontinuous charging with 532~nm light, but frequent recharging was required \cite{miyazu2011}. With near-infrared (NIR) light, KTN deflectors have been used with discontinuous charging without observing substantial photoionization, although the photoionization rate was not directly measured \cite{yagi2014}. Thus, while photoionization behavior in KTN clearly depends on wavelength, quantitative data supporting these observations are very limited.

The rate and wavelength dependencies of photoionization are determined by the defects responsible for trapping charge \cite{neamen2012, jaros1980}. In KTN, oxygen vacancies are suspected of being the dominant intrinsic defect species, though traps relevant to device operation have not been experimentally characterized \cite{yang2016, shen2012}. Materials may contain multiple trap species with distinct optical cross sections, which would produce photoionization dynamics that vary with wavelength in a complex manner. Characterizing these traps would enable prediction of photoionization behavior, guiding the design of deflectors for advanced wavelength-specific applications.

We present here a quantitative study of photoionization in KTN deflectors in the NIR-I and NIR-IIa biological imaging windows (700--1300~nm), a wavelength range commonly used for two-photon imaging. We used a two-beam interferometric method to measure changes in trapped charge density due to NIR excitation light. We report the degree of photoionization at each wavelength as a function of photon fluence. These measurements reveal multi-rate photoionization dynamics and provide a quantitative basis for predicting charge loss in KTN deflectors operating at NIR wavelengths.

\section{Experimental Methods and Model}
\begin{figure}[htbp]
    \centering
    \includegraphics[width=\linewidth]{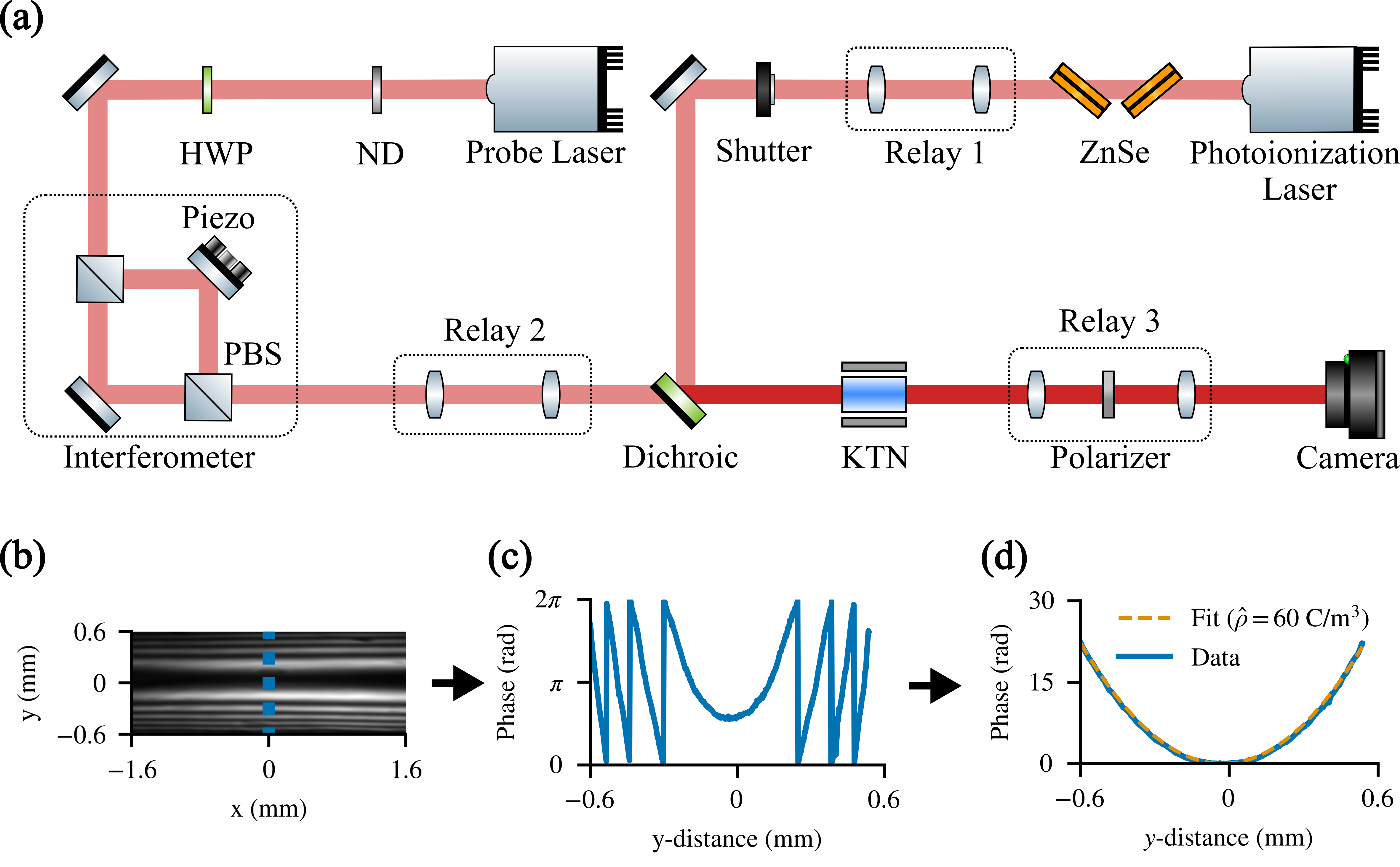}
    \caption{
    Charge density measurement method. (a) Setup: a low-power probe beam (for phase-shifting interferometry) and a gated, tunable beam (for photoionization) are combined prior to the KTN. A phase-shifting Mach-Zehnder  interferometer measures the KTN-induced differential phase. Example analysis: (b) Interferograms reflecting the phase retardation produced by the KTN are collected. The $y$-axis cross section location is indicated by a dashed blue line (c) A phase estimation algorithm converts interferograms into a wrapped phase profile along $y$. (d) A phase unwrapping algorithm is applied, and a parametric model is fit to the unwrapped phase to estimate the charge density.
    }
    \label{fig:setup}
\end{figure}

\subsection{KTN deflector}\label{sec:deflector}
Our KTN deflector employs a 3.2~mm $\times$ 1.2~mm $\times$ 4~mm ($x \times y \times z$) crystal from NTT; titanium/platinum/gold (Ti/Pt/Au) electrodes were evaporated on the $xz$ faces, and the remaining faces were polished to optical quality. In this configuration, the beam propagates along the $z$-axis and is deflected and focused in the $yz$ plane. The deflector houses a thermoelectric cooler (TEC3-2.5, Thorlabs Inc.) and a thermistor (TH10K, Thorlabs Inc.), both connected to a temperature controller (TECSource 5240, Arroyo Instruments). The temperature control system was set to keep the crystal at 28~\degree C, holding it above the cubic phase transition temperature where the Kerr effect is strongest \cite{Toyoda2014}.

\subsection{Interferometric Measurement of Charge Density}\label{sec:interferometry}
After charging with a DC voltage, a space charge is formed inside the KTN, modifying the index of refraction through the Kerr effect. Due to anisotropy of the Kerr tensor, light polarized parallel to the electric field (along $y$) is more strongly retarded than perpendicularly polarized  light (along $x$). Thus, by measuring the birefringent phase retardation, the charge density can be estimated. We built a phase-shifting Mach-Zehnder interferometer to measure the birefringent phase retardation induced by the KTN crystal \cite{chen2025, imai2014}. A linearly polarized 1064~nm laser (Femtopower 1060-fs, Fianium) served as the probe laser. The polarization was adjusted with a half-wave plate to lie at 45\degree relative to the $x$ and $y$ axes before being split into two orthogonally polarized beams by a polarizing beam splitter (PBS; PBS205, Thorlabs Inc.). In one of the arms, the beam was reflected from a mirror mounted on a piezo motion stage (NF15AP25, Thorlabs Inc.) capable of producing sub-nanometer displacements, while the other was reflected from a fixed mirror. The beams were recombined by a second PBS, after which a relay resizes the Gaussian beam to a $1/e^2$ diameter of 0.7~mm at the crystal face. The laser was adjusted to a power of \(1~\mu\text{W}\) at the KTN entrance aperture. The exit aperture of the deflector was imaged onto a CMOS camera (CS165MU, Thorlabs Inc.) using an afocal imaging system with a polarizer (LPNIRB050, Thorlabs Inc.) placed at the intermediate focal plane within the relay to create interference between the two polarizations. This configuration allows for the measurement of phase retardation between $x$- and $y$-polarized light after transmission through the crystal. The camera and the piezo mirror were computer-controlled to capture five interferogram frames with successive piston phase shifts of $\pi / 4$ between each frame. The phase retardation introduced by charging the crystal was estimated using an error-compensating phase calculation algorithm and unwrapped using a two-dimensional phase-unwrapping algorithm \cite{hariharan1987, herraez2002}. After each piezo mirror motion, the stage was allowed 2~s of settling time before the interferogram was captured. Thus, a five-frame phase estimate took 10~s.

The spatial variation of the phase retardation due to the quadratic Kerr EO effect is given by
\begin{align}
   \Delta \phi(y) &= - \frac{\kappa L_z}{2} n_0^3 (g_{11}-g_{12}) \varepsilon^2 E^2(y)
\end{align}
where $n_0$ is the index of refraction without EO effects, $g_{11}$ and $g_{12}$ are the EO coefficients for light parallel and perpendicular to $E$, respectively, $\varepsilon$ is the absolute permittivity, $L_z$ is the interaction length, and $\kappa = 2\pi/\lambda$ is the wavenumber of the probe laser \cite{imai2014}. We have fixed $x$ at the center of the crystal, as variation along this dimension is negligible due to the lack of an electric field gradient. Under conditions where the charge density $\rho$ inside the crystal is approximately uniform, we can use Gauss's law to write
\begin{align}
   E(y) = \frac{\rho}{\varepsilon}y
\end{align}
when no voltage is applied to the electrodes and $y=0$ at the crystal mid-plane. The phase retardation can then be written as
\begin{align}
   \Delta \phi(y) 
               &= - \frac{\kappa L_z}{2} n_0^3 (g_{11}-g_{12}) \rho^2 y^2 \label{eq:ktn_phase}
\end{align}
 By fitting a quadratic equation to our interferometric phase estimate, we can obtain an estimate of $|\rho|$ since all the other coefficients in Eq.~\ref{eq:ktn_phase} are known constants. We define the sign convention on the basis of electron injection polarity (negative charge).

\subsection{Measurement of Photoionization} \label{sec:meas_photo}
To characterize photoionization in the KTN crystal, we biased the electrodes at 450~V using two series-connected DC power supplies (GPR-30H100, GW Instek) to inject electrons into the crystal and fill traps. The initial charge density was measured via interferometry. We then opened a shutter (LS6, Uniblitz) to expose the device to a wavelength-tunable excitation laser (InSight X3, Spectra-Physics) for a series of \(470\) controlled exposures. After each individual exposure, the excitation laser shutter was closed and the charge density was measured following a 1~s waiting period; the charge density measurement itself takes 10~s. The cumulative exposure duration for the \(n\)th charge density measurement was given by
\begin{align}
t_n = 2.5(1.02^{\,n}-1)~~\text{s},\qquad n=0,1,\dots, 470
\label{eq:tn}
\end{align}
The exponential exposure time yields denser sampling early on, when the charge density changes most rapidly. The factor of 2.5 in Eq.~\ref{eq:tn} was chosen so that the initial exposure time was 50~ms. Including all shutter exposures, the 1~s post-exposure waiting period, and the 10~s interferometric measurement after each exposure, the total wall-clock duration of each measurement sequence is 32711~s (9.09~h).

For the sake of spatial uniformity across the crystal face, the wavelength-tunable laser was relayed to overfill the KTN aperture with a 4.4 mm $1/e^2$ diameter beam; along $y$ from the crystal center to the electrode edges, the intensity drops by 13.8\%. The laser power was set so that 100~mW was transmitted through the KTN at each wavelength. We repeated the measurement sequence for wavelengths from 700~nm to 1300~nm in 50~nm increments, always starting with a fully charged crystal.

To suppress potential two-photon processes (e.g., band-to-band absorption and two-photon photoionization), we introduced a positive group-delay dispersion (GDD) into the photoionization laser beam by placing four 5~mm ZnSe windows arranged at Brewster's angle upstream of the KTN. At each wavelength, the laser's pre-compensation was set to provide the least negative dispersion setting available. 

Assuming a Gaussian spectrum and purely quadratic spectral phase, we calculated the worst-case pulse width and corresponding peak intensity we would expect for these beam properties. We conservatively assumed that the transform-limited pulse FWHM was 100 fs, and extrapolated the GDD at the laser output from manufacturer specifications. We also calculated the photon fluence $\mathcal{F}_{\gamma,1/e}$ (where $\gamma$ denotes photon) required to reduce the trapped-charge density by $1/e$, assuming the degenerate two-photon absorption cross section to be 1~GM (Goeppert-Mayer unit; $1~\text{GM} \equiv 10^{-50}~\mathrm{cm^4\,s\,photon^{-1}}$).  In the NIR, the degenerate two-photon absorption cross section for nitrogen vacancies in diamond has been reported to be $<1$~GM for transitions between defect states \cite{ivaylo2013}. Given the larger energy separation for trap ionization in KTN, 1 GM is a conservative upper bound for our estimates. Table~\ref{tab:znse_pulse_peak} shows the results of these calculations; $\mathcal{F}_{\gamma,1/e}$ serves as an upper bound on the two-photon photoionization rate and can be compared to the measured photoionization data.

\begin{table*}[htbp]
\centering
\caption{GDD-broadened pulse duration and corresponding peak quantities
for a 100~mW transmitted average power at 80~MHz. The beam is Gaussian
with $D_{1/e^2}=4.4~\text{mm}$ overfilling a $3.2~\text{mm}\times 1.2~\text{mm}$
aperture; Input beam was assumed to have a FWHM of 100 fs. $I_{0,\mathrm{peak}}$ is the on-axis (center) peak intensity. $\tau_{\mathrm{out}}$ is the pulse FWHM following four 5~mm ZnSe windows.}
\label{tab:znse_pulse_peak}
\begin{adjustbox}{width=\columnwidth,center}
\begin{tabular}{r r r r r r r}
\toprule
$\lambda$ (nm) & $\mathrm{GDD}_{\mathrm{pre}}$ (fs$^2$) & $\mathrm{GDD}_{\mathrm{ZnSe}}$ (fs$^2$) & $\mathrm{GDD}_{\mathrm{tot}}$ (fs$^2$) & $\tau_{\mathrm{out}}$ (fs) & $I_{0}$ (W/cm$^2$) & $\mathcal{F}_{\gamma,1/e}$ (photons/cm$^2$) \\
\midrule
$700$ & $-10\,000$ & $29\,968$ & $19\,968$ & $562.6$ & $7.75\times10^{4}$ & $5.18\times10^{26}$ \\
$750$ & $-5\,000$ & $25\,366$ & $20\,366$ & $573.5$ & $7.61\times10^{4}$ & $4.92\times10^{26}$ \\
$800$ & $0$ & $22\,103$ & $22\,103$ & $620.9$ & $7.02\times10^{4}$ & $5.00\times10^{26}$ \\
$850$ & $0$ & $19\,658$ & $19\,658$ & $554.1$ & $7.87\times10^{4}$ & $4.20\times10^{26}$ \\
$900$ & $0$ & $17\,750$ & $17\,750$ & $502.2$ & $8.69\times10^{4}$ & $3.59\times10^{26}$ \\
$950$ & $0$ & $16\,214$ & $16\,214$ & $460.5$ & $9.47\times10^{4}$ & $3.12\times10^{26}$ \\
$1000$ & $0$ & $14\,948$ & $14\,948$ & $426.3$ & $1.02\times10^{5}$ & $2.75\times10^{26}$ \\
$1050$ & $0$ & $13\,883$ & $13\,883$ & $397.7$ & $1.10\times10^{5}$ & $2.44\times10^{26}$ \\
$1100$ & $-600$ & $12\,972$ & $12\,372$ & $357.3$ & $1.22\times10^{5}$ & $2.09\times10^{26}$ \\
$1150$ & $-1\,200$ & $12\,182$ & $10\,982$ & $320.5$ & $1.36\times10^{5}$ & $1.79\times10^{26}$ \\
$1200$ & $-1\,800$ & $11\,490$ & $9\,690$ & $286.7$ & $1.52\times10^{5}$ & $1.54\times10^{26}$ \\
$1250$ & $-2\,400$ & $10\,878$ & $8\,478$ & $255.4$ & $1.71\times10^{5}$ & $1.32\times10^{26}$ \\
$1300$ & $-3\,000$ & $10\,331$ & $7\,331$ & $226.5$ & $1.93\times10^{5}$ & $1.12\times10^{26}$ \\
\bottomrule
\end{tabular}
\end{adjustbox}
\end{table*}

\subsection{Model of Photoionization Dynamics} \label{sec:photo_dynamic}

We modeled the photoionization dynamics in the KTN deflector using a drift-diffusion-Poisson system with electrostatic coupling. We model the system one-dimensionally along the charging dimension $y$. After DC charging, we assumed that the crystal contains \(M\) distinct trap species, each with density \(N_{T,i}\). We allowed each trap to contain between zero and \(K_i\) electrons to model defects with multiple charge states \cite{watkins1984, theis1990}. The portion of \(i\)-species traps containing \(k\) electrons is defined as \(f_{i,k}(y,t)\), subject to
\begin{equation}
\sum_{k=0}^{K_i} f_{i,k} (y,t) = 1
\label{eq:occupancy}
\end{equation}
Let \(n(y,t)\) denote the free-electron density. The total charge density inside the crystal is then
\begin{equation}
\rho(y,t) = -q\!\left[n(y,t) + \sum_{i=1}^M N_{T,i} \sum_{k=1}^{K_i} kf_{i,k}(y,t)\right]
\label{eq:rho}
\end{equation}
where \(q\) is the magnitude of the electron charge. Under illumination with photon flux \(\Phi_\gamma(y,t)\), the evolution of the trap occupancies is
\begin{align}
\frac{\partial f_{i,k}(y,t)}{\partial t}
  &= \sigma_{i,k+1} \Phi_\gamma(y,t)\,f_{i,k+1}(y,t)
   - \sigma_{i,k} \Phi_\gamma(y,t)\,f_{i,k}(y,t) \notag \\
  &\quad + C_{i,k-1}\,n(y,t)\,f_{i,k-1}(y,t)
   - C_{i,k}\,n(y,t)\,f_{i,k}(y,t)
\label{eq:dfdt}
\end{align}
In Eq.~\ref{eq:dfdt}, \(\sigma_{i,k}\) is the photoionization cross section and \(C_{i,k}\) is the capture cross section for trap \(i\) when holding \(k\) electrons. By definition, \(C_{i,K_i}\) and \(\sigma_{i,0}\) must be zero. The free-electron density evolves according to the electron continuity equation
\begin{equation}
\frac{\partial n(y,t)}{\partial t}
  = -\frac{1}{q}\frac{\partial J_n(y,t)}{\partial y}
    + \sum_{i=1}^M N_{T,i} \sum_{k=1}^{K_i}
      \left[\sigma_{i,k} \Phi_\gamma(y,t)\,f_{i,k}(y,t)
            - C_{i,k-1}\,n(y,t)\,f_{i,k-1}(y,t)\right]
\label{eq:dndt}
\end{equation}
where the electron current density is
\begin{equation}
J_n(y,t)
  = q\mu\,n(y,t)\,E(y,t)
    - qD\,\frac{\partial n(y,t)}{\partial y}
\label{eq:j}
\end{equation}
Here, \(\mu\) is the electron mobility and \(D\) is the diffusion coefficient. The boundary condition for the electric potential is given by
\begin{equation}
  V(L/2,t) - V(-L/2,t) = 0
  \label{eq:bc_phi}
\end{equation}
where $L$ is the electrode gap length. We choose to fix $V(L/2,t) = 0$. Equation~\ref{eq:bc_phi} corresponds to the electrodes being shorted after removal of the charging voltage. Immediately after the charging voltage is removed, all traps are filled and the free-carrier concentration is negligible, so
\begin{equation}
\rho(y,0) = -q\sum_{i=1}^M N_{T,i} K_i
\label{eq:rho0}
\end{equation}

\subsubsection{Charge Density Uniformity} 
\label{sec:uniform}

Given uniform initial conditions and uniform illumination $\Phi_\gamma(y,t)=\Phi_\gamma(t)$, the occupancy factors $f_{i,k}$ remain uniform inside the crystal during photoionization under the following uniform conditions:

\begin{equation}
\mathcal{U} = \big\{\, n(y,t) = N(t), \;\; f_{i,k}(y,t)=F_{i,k}(t)\, \big\}.
\label{eq:U}
\end{equation}
the charge density is also spatially uniform, and we denote it by $\rho(t)$. Gauss's law can be written as
\begin{equation}
\varepsilon\,\frac{\partial E(y,t)}{\partial y} = \rho(t)
  = -q\!\left[N(t) + \sum_{i=1}^M N_{T,i} \sum_{k=1}^{K_i} k\,F_{i,k}(t)\right].
\label{eq:gauss}
\end{equation}
Eq.~\ref{eq:gauss} implies that \(E(y,t)\) is linear in \(y\) with slope determined by the charge density. Since $\partial_y n = 0$ and $\partial_y E$ is constant under uniform charge density, the divergence $\partial_y J_n$ is independent of $y$. We can then write the system of ordinary differential equations as
\begin{align}
\frac{dF_{i,k}(t)}{dt}
  &= \sigma_{i,k+1} \Phi_\gamma(t)\,F_{i,k+1}(t)
   - \sigma_{i,k} \Phi_\gamma(t)\,F_{i,k}(t) \notag \\
  &\quad + C_{i,k-1}\,N(t)\,F_{i,k-1}(t)
   - C_{i,k}\,N(t)\,F_{i,k}(t),
\label{eq:Fik_ODE}\\[3pt]
\frac{dN(t)}{dt}
  &= -\frac{\mu}{\varepsilon}\,N(t)\,\rho(t)
     - \sum_{i=1}^M N_{T,i} \sum_{k=1}^{K_i} k\,\frac{dF_{i,k}(t)}{dt}.
\label{eq:N_ODE}
\end{align}
Since the partial differential equations admit spatially uniform solutions consistent with the initial and boundary conditions \eqref{eq:rho0}, the spatially uniform subspace \eqref{eq:U} is invariant. Thus, under uniform illumination, the trap occupancies and total charge density remain uniform throughout the crystal and the dynamics reduce to a lumped ODE model. The model is a bulk approximation; we assume thin boundary layers near the interfaces accommodate the carrier-flux boundary conditions and can be neglected.

With our experimental technique, we could only directly observe the total charge density $\rho(t)$, which includes the free charge. However, because of the time required to perform the interferometric measurement, we restricted ourselves to measuring the charge density in a quasi-steady state where all free charge has either been trapped or left the crystal. We note that, although we did not directly capture the dynamic behavior of $N(t)$, it is nonzero following photoionization. 

The nonlinear dynamics of Eqs.~\ref{eq:Fik_ODE}--\ref{eq:N_ODE} combined with the non-observability of $N(t)$ in our experimental setup made it impractical to fit the data to the complete model. Recapture rates in KTN have not been reported. In the absence of independent measurements of the rate of recapture, it would not be possible to reliably distinguish between the existence of multiple traps or a single trap with complex recapture dynamics. To help direct future investigations of the trap properties in KTN, we sought to rule out the simplest possible defect: a single trap with negligible recapture.  Thus, we set the recapture rate ($C_{i,k}$) to zero. Furthermore, density functional theory calculations suggest a doubly ionized oxygen vacancy ($V_O^{+2}$) may be the dominant electron trap species in KTN \cite{shen2012,yang2016}. Based on this, we modeled an oxygen vacancy as a single trap ($M=1$) that can hold two electrons ($K=2$). The no-recapture assumption allowed us to decouple the trap occupancy rate equations and linearize the system. With no recapture, the trap dynamics depend only on the cumulative photon fluence
\begin{equation}
\mathcal{F}_\gamma(t) = \int_0^t \Phi_\gamma(t')\, dt'.
\label{eq:fluence_def}
\end{equation}
With initial conditions $F_2(0) = 1$, $F_1(0) = 0$, $F_0(0) = 0$ corresponding to a fully charged crystal, the trap occupancies can be explicitly written as:
\begin{align}
F_2(\mathcal{F}_\gamma) &= e^{-\sigma_2 \mathcal{F}_\gamma}, \label{eq:F2_sol}\\[8pt]
F_1(\mathcal{F}_\gamma) &= 
\begin{cases}
\displaystyle \frac{\sigma_2}{\sigma_1 - \sigma_2}\left(e^{-\sigma_2 \mathcal{F}_\gamma} - e^{-\sigma_1 \mathcal{F}_\gamma}\right), & \sigma_1 \neq \sigma_2, \\[12pt]
\sigma_1 \mathcal{F}_\gamma \, e^{-\sigma_1 \mathcal{F}_\gamma}, & \sigma_1 = \sigma_2,
\end{cases} \label{eq:F1_sol}\\[8pt]
F_0(\mathcal{F}_\gamma) &= 1 - F_1(\mathcal{F}_\gamma) - F_2(\mathcal{F}_\gamma). \label{eq:F0_sol}
\end{align}
In Eqs.~\ref{eq:F2_sol}--\ref{eq:F0_sol}, $\sigma_1$ and $\sigma_2$ are wavelength-dependent. While $N(t)$ is generally nonzero immediately following photoionization, our measurements are performed after sufficient delay that the free carriers no longer contribute to the observed charge density. This is supported by the observation that immediately following an exposure, the interferogram changes rapidly for a brief period (<1 s) before stabilizing. In this case, we can write the charge density as
\begin{equation}
\rho(\mathcal{F}_\gamma) = -qN_T \left[F_1(\mathcal{F}_\gamma)+2F_2(\mathcal{F}_\gamma)\right],
\label{eq:P_phi}
\end{equation}
which is a more tenable form for fitting to our experimental data.

\section{Results and Discussion}
\begin{figure}[htbp]
    \centering
    \hspace*{-5mm}\includegraphics[]{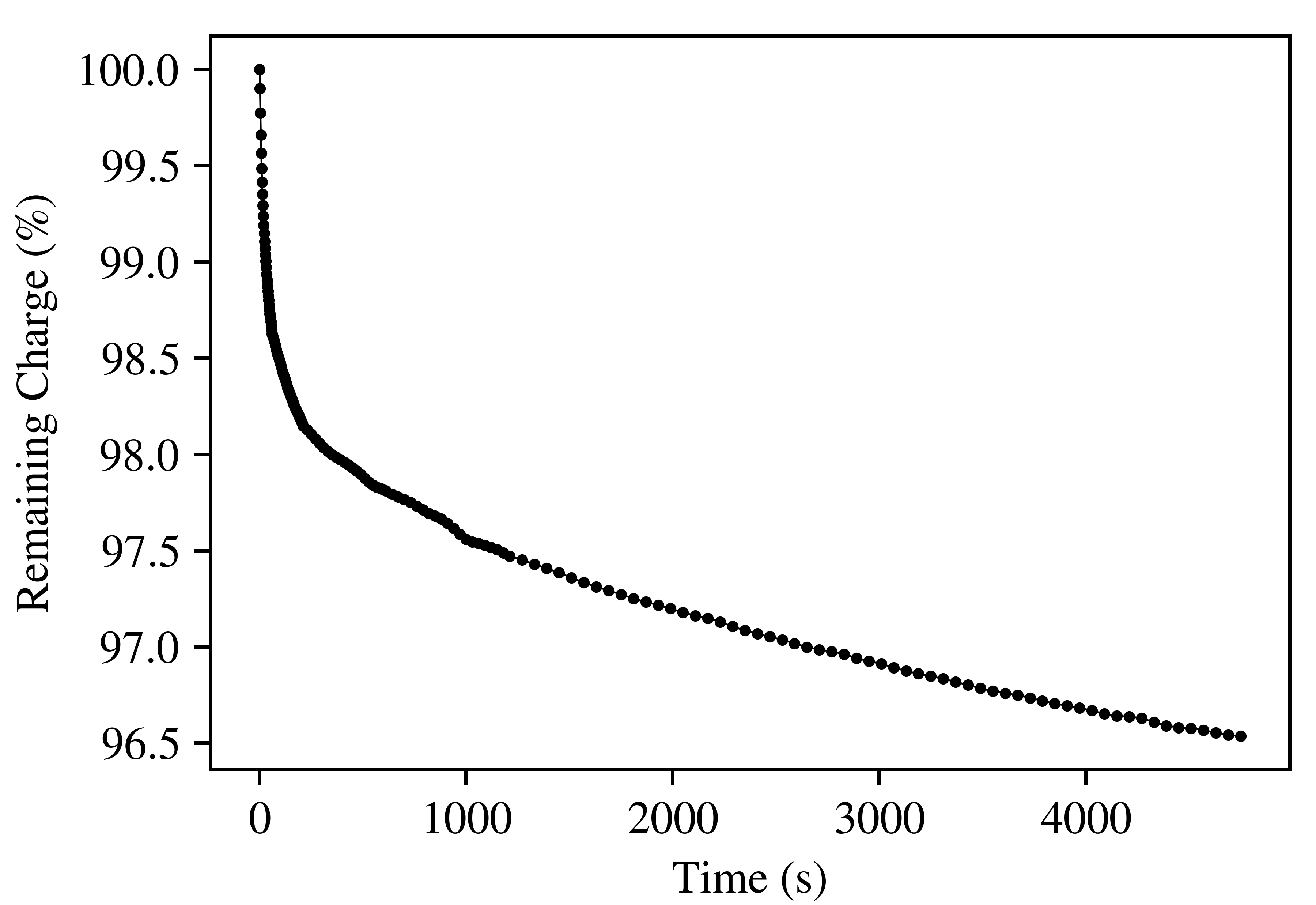}
    \caption{
    Remaining charge (percentage of initial charge) vs. time for a KTN crystal in the absence of a scanning laser. KTN charged to 450~V and operated at 28~\degree C. 
    }
    \label{fig:dark_depl}
\end{figure}

Our initial qualitative observations showed that at operating temperature (28~\degree C), with no illumination of the KTN, the trapped electrons were stable. To quantify the rate at which charge is lost in the absence of the excitation laser, we performed time-resolved measurements of the charge density, with only the 1~$\mu$W probe beam present. The experimental setup is shown in Fig.~\ref{fig:setup}. The results of this experiment are shown in Fig.~\ref{fig:dark_depl}. Following removal of the charging voltage, we observed that over 96\% of the initial charge density remained after 80 minutes, with half of the loss in charge taking place within the first 6 minutes.

With an understanding of the stability of the trapped charge in the absence of photoionization light, we performed a set of experiments to characterize the photoionization due to light from 700~nm to 1300~nm in 50~nm steps. This wavelength range covers most of the common application wavelengths used in two-photon excitation microscopy. The experiment involved sequentially exposing a charged KTN crystal to the photoionization laser for a brief period and then measuring the remaining charge density. 

\begin{figure}[htbp]
    \centering
    \includegraphics[width=\linewidth]{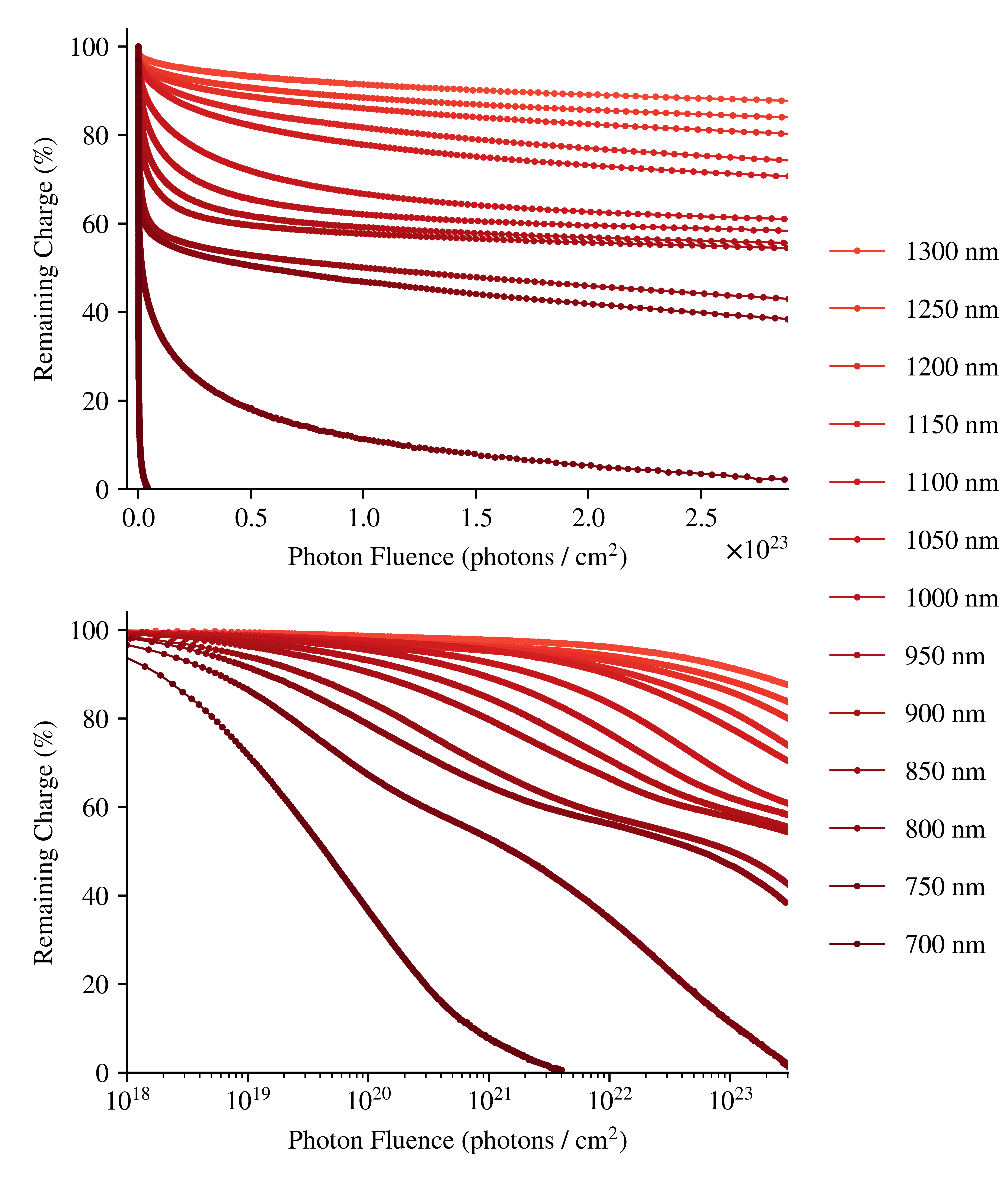}
    \caption{
    Wavelength-dependent photoionization in KTN from 700~nm to 1300~nm in 50~nm increments. Remaining charge (percentage of initial charge) vs. photon fluence, shown on linear (top) and logarithmic (bottom) $x$-axes. 
    }
    \label{fig:photoionization}
\end{figure}

The results of the photoionization experiment are shown in Figure~\ref{fig:photoionization}. The remaining charge density is plotted against photon fluence rather than time, since photoionization is driven by the number of photons incident on the device. The data are plotted on linear and logarithmic $x$-axes to visualize the decaying exponential behavior that spans multiple timescales. On the lower $x$-axis in Figure~\ref{fig:photoionization}, markers are used to denote the photon fluence that would result from three example beams, described in the caption. These markers are provided to help relate typical beam parameters to photon fluence.

When we compare charge density decay in the absence of illumination (Fig.~\ref{fig:dark_depl}) to decay under 1300~nm illumination in terms of wall-clock time since the start of each experiment, the difference is small. The final measurement in the dark sequence was acquired at $t=4750$~s at which point 96.5\% of the initial charge remained. Interpolating the 1300~nm dataset to the same wall-clock time yields 96.3\% remaining charge. However, this near-concordance does not mean that 1300~nm illumination produces no photoionization. Differences may become apparent across longer simultaneous observation windows or with higher average photon flux. 

Within the range of the experimental photon fluence, after $\approx$ 9 hours of exposure to 100 mW only the 700~nm beam was seen to fully deplete the crystal of trapped charge. As expected from theoretical models, the photoionization rate decreased dramatically as the wavelength was increased. The fluence sufficient to fully deplete the crystal at 700~nm releases less than 5\% of the initially trapped charge at 1300~nm. The data in Fig.~\ref{fig:photoionization} also reveal a multi-exponential photoionization response. At 900~nm, the decay is initially sharp but then appears to asymptote around 60\% charge density. Beyond $10^{23}$ photons / cm$^2$ the decay rate is actually higher at 1100~nm than it is at 900~nm. 

For the 700~nm and 750~nm curve in Fig.~\ref{fig:photoionization}, we can compare the photon fluence required for the charge density to reach $1/e$ (36.8\%) of its initial value, $\mathcal{F}_{\gamma,1/e}$, to our calculations in Table ~\ref{tab:znse_pulse_peak}. Despite assuming a high two-photon cross section of $1$~GM, the experimentally observed $\mathcal{F}_{\gamma,1/e}$ is multiple orders of magnitude smaller than our calculations. This suggests that a two-photon process is unlikely to be dominant at these wavelengths. Above 800~nm, the decay of the normalized charge density is incomplete and has a stronger multi-exponential character. The interpretation in this regime is less clear, although the partial depletion that occurs appears to proceed at a rate faster than predicted in Table ~\ref{tab:znse_pulse_peak}. 

The experimental technique developed here has many similarities to the photo-stimulated discharge (PSD) method used to evaluate charge-trap energy levels \cite{boudou2018,mendoza-lopez2022}. PSD is most often used in polymer dielectrics, but it is applicable to many transparent or semitransparent materials \cite{zheng2017, teyssedre_2021}. In its most common form, PSD involves charging a sample via electrodes and monitoring the electrode current during illumination. The excitation wavelength is typically swept continuously from low to high energy. As the photon energy approaches the energy required to photoionize a given defect population, an increase in current is observed. Plotting the measured current versus excitation wavelength yields a PSD spectrum. Some studies also collect time-dependent current traces at discrete wavelengths and use the area under each curve as a proxy for the trapped charge associated with defect levels accessible within the excitation spectrum \cite{zhu2009}.

The main difference between our experiment and PSD studies is that we optically measure the residual charge after a photon fluence dose as opposed to electric current during exposure. Fundamentally, the quantities measured by both techniques reflect the photoionization rate as a function of wavelength. PSD studies differ in how they interpret their spectra or time traces, with some claiming to identify specific defect energies and others choosing not to take a specific interpretation of the data. The photoionization response from a single trap species can be spectrally complex, and the effects of recapture, multi-electron traps, photoionization from distinct trap species, and potential multi-photon photoionization make the measured photoionization responses more difficult to interpret. Thus, cautious use of a model with the ability to account for these effects is warranted. 

We developed a model to aid in the interpretation of the photoionization results shown in Fig.~\ref{fig:photoionization}. We use a simplified form of this model to test the hypothesis that trap dynamics in KTN are driven by a single species of oxygen vacancy (Eqs.~\ref{eq:F1_sol}--\ref{eq:P_phi}). The simplified model includes a single defect species capable of trapping up to two electrons. Recapture of electrons by KTN defects following photoionization may be significant. However, to linearize the model and avoid parameter degeneracy, recapture was assumed to be negligible ($C_{i,k}\approx0$). The measured charge density values $\rho(\mathcal{F}_\gamma)$ were normalized by the initial charge density $\rho_0$, which was assumed to be a good approximation of $-qN_T$. The optical cross section parameters $\sigma_1$ and $\sigma_2$ were fit independently at each wavelength with no assumed spectral form using a nonlinear least squares approach.

\begin{figure}[htbp]
    \centering
    \hspace*{-10mm}\includegraphics[]{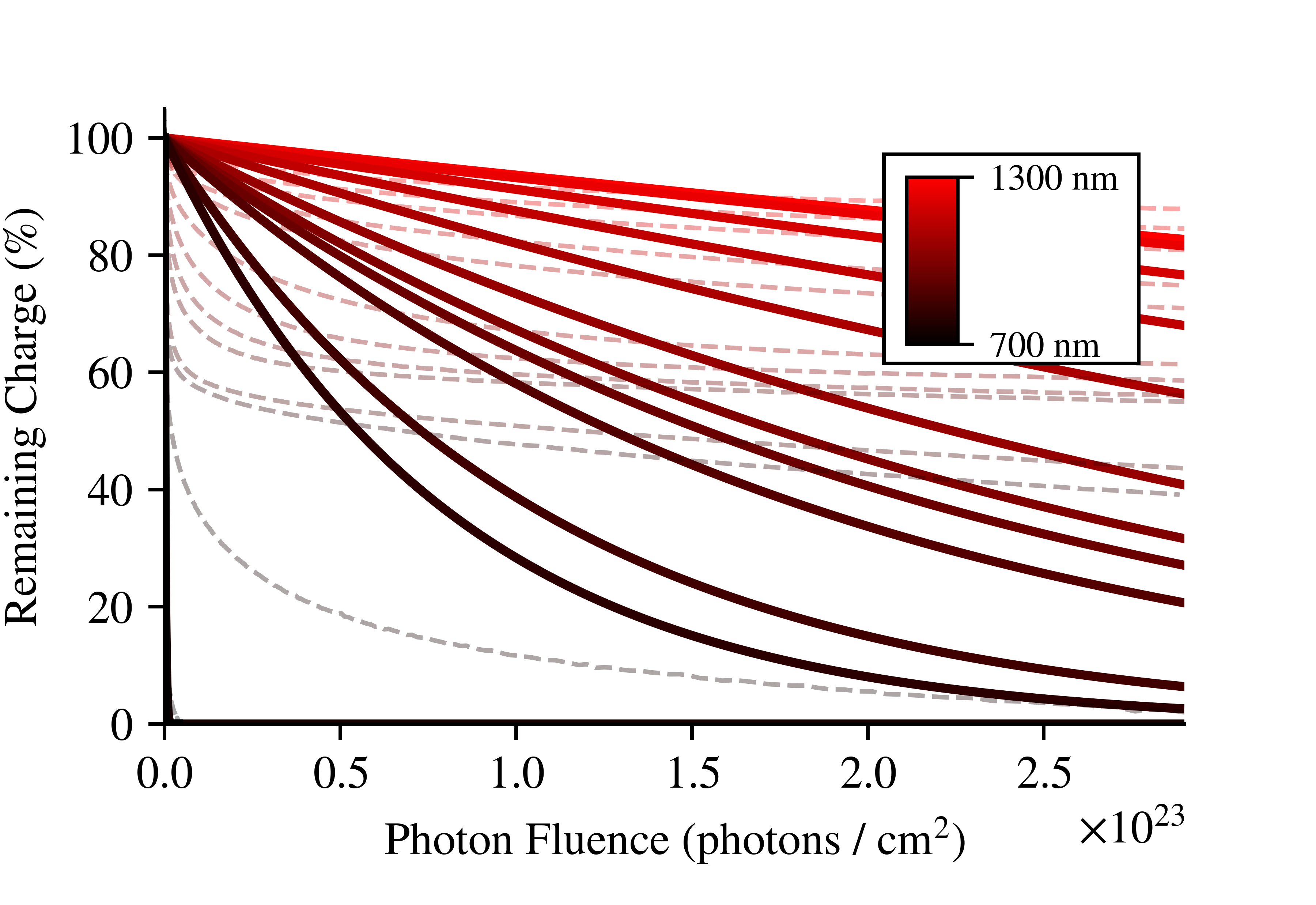}
    \caption{
    Remaining charge (percentage of initial charge) vs. photon fluence at wavelengths from 700~nm to 1300~nm in 50~nm increments. Color indicates wavelength (see colorbar). Dashed lines show experimental data and solid lines show the best-fit (least-squares) model (Eqs.~\ref{eq:F1_sol}--\ref{eq:P_phi}). $\sigma_1$ and $\sigma_2$ are fit independently at each wavelength. 
    }
    \label{fig:fit}
\end{figure}

The fitting results are shown in Fig.~\ref{fig:fit}, where the remaining charge density is plotted as a function of cumulative photon fluence (Eq.~\ref{eq:fluence_def}) for each wavelength. Large qualitative differences in behavior can be seen between the best-fit models and the experimental data. As a quantitative measure of goodness-of-fit, the normalized root mean square error (NRMSE) was computed for each wavelength on the normalized charge. Across wavelength, the two-electron, no-recapture single-trap model yields a median NRMSE of 0.136 (Interquartile Range: IQR = 0.138, Q1--Q3: 0.053--0.191) on normalized charge density, indicating substantial systematic mismatch.

There are a number of possible explanations for the failure of the model to explain the data. Recapture of photoionized electrons can nonlinearize the system and change the solution structure. Alternatively, local chemistry around the $V_O^{+2}$ defect (relative number of nearest-neighbor B-site Ta/Nb) or local strains could strongly affect trap properties, leading to a spectrum of trap properties. It is also possible that there are additional species of electron traps in KTN besides $V_O^{+2}$ which are relevant to device operation. Determining the molecular origin of the electron traps in KTN would likely require the application of experimental techniques such electron paramagnetic resonance (EPR) or deep-level transient spectroscopy (DLTS) \cite{lu2019, dobaczewski2004}.

\section{Conclusion}
We present systematic quantitative measurements of photoionization in KTN electro-optical deflectors across infrared wavelengths covering the NIR-I and NIR-IIa biological imaging windows. We observed that photoionization rates, even at the lowest near-infrared wavelengths, were much slower than the rates reported in the visible range \cite{yagi2014}. These rates decreased by many orders of magnitude with increasing wavelength across the NIR range. This suggests that, for NIR applications, KTN-based deflectors may not require a bias voltage for continuous recharging. Rather, performance could be maintained via brief interruptions of the scan for recharging. This data provides quantitative guidance for prediction charge-loss-induced performance degradation, selection of wavelengths for performance-sensitive applications, and determination of charge-scan duty cycles for NIR applications. These insights will be particularly critical for the implementation of advanced methods to increase the range of electro-optical deflection via relayed optical loops \cite{jayakumar2025}, wherein the application of a substantial bias voltage severely degrades performance.

The multi-exponential nature of the photoionization decay curves suggests the existence of multiple trap species. When similar behavior has been observed in PSD studies\cite{boudou2018,mendoza-lopez2022}, it is typically interpreted as evidence of multiple trap species. We quantitatively demonstrate that, in absence of recapture, our measurements are not consistent with a single defect, even a defect with two-electron vacancy. However, we did not independently measure recapture, thus a single defect species cannot be completely ruled out. Future work can build on these results to obtain more detailed characterizations of defect properties in KTN.The precise identification of multiple trap species may facilitate advanced wavelength-based methods for increasing or sculpting the space-charge in KTN-based deflectors, improving the performance of the next generation of KTN-based deflectors.

\begin{backmatter}
\bmsection{Funding}National Institutes of Health (RF1NS128658). Fabrication at Minnesota Nano Center: National Science Foundation (ECCS-2025124).
\bmsection{Acknowledgment}Machining advice and services were provided by the College of Science and Engineering Shop at the University of Minnesota. Data for charge depletion in the absence of a scanning laser (Fig. 3)  was previously presented in the Master's thesis of SS \cite{stanek2024}.  
\bmsection{Disclosures} HJ, CW, AK: Arc Photonics (I).
\bmsection{Data Availability Statement}Data and analysis are available at www.github.com/kerlin-lab/Stanek\_et\_al\_2026.
\end{backmatter}

\bibliography{refs.bib}

@article{shen2012,
title = {Structural and electronic properties of oxygen vacancy in lead-free KTa1-xNbxO3: Comparative first-principles calculations},
journal = {Computational Materials Science},
volume = {65},
pages = {193-196},
year = {2012},
issn = {0927-0256},
doi = {https://doi.org/10.1016/j.commatsci.2012.07.024},
url = {https://www.sciencedirect.com/science/article/pii/S0927025612004363},
author = {Yanqing Shen and Zhongxiang Zhou}
}

@article{yang2016,
  author  = {Yang, Wei and Wang, Lei and Lin, Jing and others},
  title   = {Effect of {A}-Site and Oxygen Vacancies on the Structural and Electronic Properties of Lead-Free {KTa$_{0.5}$Nb$_{0.5}$O$_3$} Crystal},
  journal = {Journal of Electronic Materials},
  volume  = {45},
  number  = {},
  pages   = {3726--3733},
  year    = {2016},
  doi     = {10.1007/s11664-016-4477-6}
}

@article{imai2014,
  title={Measurement of charge density distributions in KTa1-xNbxO3 optical beam deflectors},
  author={Imai, Tadayuki and Miyazu, Jun and Kobayashi, Junya},
  journal={Optical Materials Express},
  volume={4},
  number={5},
  pages={976--981},
  year={2014},
  publisher={Optical Society of America}
}

@article{chen2025,
title = {Measurement of space charge density distributions and dielectric resonance enhancement of beam deflection properties of KTN crystal},
journal = {Journal of Materiomics},
volume = {11},
number = {3},
pages = {100902},
year = {2025},
issn = {2352-8478},
doi = {https://doi.org/10.1016/j.jmat.2024.04.017},
url = {https://www.sciencedirect.com/science/article/pii/S235284782400128X},
author = {Pan Chen and Wendie Chen and Shuo Zhang and Jianwei Zhang and Jianxing Shen and Bing Liu and Xuping Wang}
}

@article{theis1990,
  author  = {Theis, T. N.},
  title   = {The DX Center: How Complicated can a Point Defect Be?},
  journal = {MRS Online Proceedings Library},
  volume  = {209},
  pages   = {367--378},
  year    = {1990},
  month   = oct,
  doi     = {10.1557/PROC-209-367},
  url     = {https://doi.org/10.1557/PROC-209-367}
}

@Inbook{watkins1984,
author="Watkins, George D.",
title="Negative-U properties for defects in solids",
bookTitle="Advances in Solid State Physics: Plenary Lectures of the 48th Annual Meeting of the German Physical Society (DPG) and of the Divisions ``Semiconductor Physics'' ``Metal Physics'' ``Low Temperature Physics'' ``Thermodynamics and Statistical Physics'' ``Thin Films'' ``Surface Physics'' ``Magnetism'' ``Physics of Polymers'' ``Molecular Physics'' M{\"u}nster, March 12{\ldots}17, 1984",
year="1984",
publisher="Springer Berlin Heidelberg",
address="Berlin, Heidelberg",
pages="163--189",
isbn="978-3-540-75374-2",
doi="10.1007/BFb0107450",
url="https://doi.org/10.1007/BFb0107450"
}

@article{zhu2018,
    author = {Zhu, Wenbin and Chao, Ju-Hung and Chen, Chang-Jiang and Shang, Annan and Lee, Yun Goo and Yin, Shizhuo and Dubinskii, Mark and Hoffman, Robert C.},
    title = {Photon excitation enabled large aperture space-charge-controlled potassium tantalate niobate (KTN) beam deflector},
    journal = {Applied Physics Letters},
    volume = {112},
    number = {13},
    pages = {132901},
    year = {2018},
    month = {03},
    issn = {0003-6951},
    doi = {10.1063/1.5021958},
    url = {https://doi.org/10.1063/1.5021958},
    eprint = {https://pubs.aip.org/aip/apl/article-pdf/doi/10.1063/1.5021958/14509980/132901_1_online.pdf},
}

@article{nakamura2008,
    author = {Nakamura, Koichiro and Miyazu, Jun and Sasaki, Yuzo and Imai, Tadayuki and Sasaura, Masahiro and Fujiura, Kazuo},
    title = {Space-charge-controlled electro-optic effect: Optical beam deflection by electro-optic effect and space-charge-controlled electrical conduction},
    journal = {Journal of Applied Physics},
    volume = {104},
    number = {1},
    pages = {013105},
    year = {2008},
    month = {07},
    issn = {0021-8979},
    doi = {10.1063/1.2949394},
    url = {https://doi.org/10.1063/1.2949394},
    eprint = {https://pubs.aip.org/aip/jap/article-pdf/doi/10.1063/1.2949394/15014079/013105_1_online.pdf},
}

@misc{NTTAT_KTN_deflector,
  title        = {KTN Optical Deflector},
  howpublished = {\url{https://keytech.ntt-at.com/en/ktn_crystal/prd_2049.html}},
  organization = {NTT Advanced Technology Corporation},
  note         = {Accessed: 2025-12-19},
  year         = {n.d.},
}

@article{miyazu2011,
doi = {10.1143/APEX.4.111501},
url = {https://doi.org/10.1143/APEX.4.111501},
year = {2011},
month = {oct},
publisher = {},
volume = {4},
number = {11},
pages = {111501},
author = {Miyazu, Jun and Imai, Tadayuki and Toyoda, Seiji and Sasaura, Masahiro and Yagi, Shogo and Kato, Kazutoshi and Sasaki, Yuzo and Fujiura, Kazuo},
title = {New Beam Scanning Model for High-Speed Operation Using KTa1-xNbxO3 Crystals},
journal = {Applied Physics Express}
}

@article{wang2023,
title = {Potassium tantalate niobate crystals: Efficient quadratic electro-optic materials and their laser modulation technology},
journal = {Journal of Materiomics},
volume = {9},
number = {5},
pages = {838-854},
year = {2023},
issn = {2352-8478},
doi = {https://doi.org/10.1016/j.jmat.2023.02.006},
url = {https://www.sciencedirect.com/science/article/pii/S2352847823000370},
author = {Xuping Wang and Xinguo Mao and Pan Chen and Qian Du and Yuguo Yang and Panyu Qiao and Shaodong Zhang and Zhijian Li and Rui Zhang and Bing Liu and Jiyang Wang}
}

@article{farinella2024,
author = {Deano M. Farinella and Samuel Stanek and Harishankar Jayakumar and Zachary L. Newman and Jacob Gable and James Leger and Aaron Kerlin},
title = {{Two-dimensional electro-optical multiphoton microscopy}},
volume = {11},
journal = {Neurophotonics},
number = {2},
publisher = {SPIE},
pages = {025005},
year = {2024},
doi = {10.1117/1.NPh.11.2.025005},
URL = {https://doi.org/10.1117/1.NPh.11.2.025005}
}

@inproceedings{yin2021,
author = {Yun Goo Lee and Annan Shang Jr. and Ruijia Liu and Wei Zhang and Shizhuo Yin},
title = {{2-dimensional KTN deflector enabled large capacity time division multiplexing based beam combining}},
volume = {11826},
booktitle = {Photonic Fiber and Crystal Devices: Advances in Materials and Innovations in Device Applications XV},
editor = {Shizhuo Yin and Ruyan Guo},
organization = {International Society for Optics and Photonics},
publisher = {SPIE},
pages = {118260B},
year = {2021},
doi = {10.1117/12.2596904},
URL = {https://doi.org/10.1117/12.2596904}
}

@INPROCEEDINGS{jimenez2023,
       author = {{Mart{\'\i}nez Jim{\'e}nez}, Alejandro and {Grelet}, Sacha and {Bowen Montague}, Patrick and {Bradu}, Adrian and {Podoleanu}, Adrian},
        title = "{Dual ultrahigh speed swept-source and time domain optical coherence tomography system using a time stretch laser and a KTN deflector}",
    booktitle = {Society of Photo-Optical Instrumentation Engineers (SPIE) Conference Series},
         year = 2023,
       editor = {{Vakoc}, Benjamin J. and {Wojtkowski}, Maciej and {Yasuno}, Yoshiaki},
       series = {Society of Photo-Optical Instrumentation Engineers (SPIE) Conference Series},
       volume = {12632},
        month = aug,
          eid = {126320B},
        pages = {126320B},
          doi = {10.1117/12.2670440},
       adsurl = {https://ui.adsabs.harvard.edu/abs/2023SPIE12632E..0BM}
}

@article{okabe2013,
  author  = {Okabe, Yuichi and Sasaki, Yuzo and Ueno, Masahiro and Sakamoto, Takashi and Toyoda, Seiji and Kobayashi, Junya and Ohmi, Masato},
  title   = {High-Speed Optical Coherence Tomography System Using a 200-kHz Swept Light Source with a {KTN} Deflector},
  journal = {Optics and Photonics Journal},
  volume  = {3},
  pages   = {190--193},
  year    = {2013},
  doi     = {10.4236/opj.2013.32031},
}

@article{damodaran2018,
author = {Vani Damodaran and Nilesh J. Vasa and R. Sarathi},
journal = {Appl. Opt.},
number = {11},
pages = {2915--2922},
publisher = {Optica Publishing Group},
title = {KTN-based high-speed axial and lateral scanning technique for an optical coherence tomography system and application to dental imaging},
volume = {57},
month = {Apr},
year = {2018},
url = {https://opg.optica.org/ao/abstract.cfm?URI=ao-57-11-2915},
doi = {10.1364/AO.57.002915},
}

@INPROCEEDINGS{sakamoto2014,
  author={Sakamoto, Takashi and Toyoda, Seiji and Ueno, Masahiro and Kobayashi, Junya},
  booktitle={IEEE CPMT Symposium Japan 2014}, 
  title={High-speed optical beam scanning using KTN crystal}, 
  year={2014},
  volume={},
  number={},
  pages={173-176},
  doi={10.1109/ICSJ.2014.7009638}
}

@article{Toyoda2014,
    author = {Toyoda, Seiji and Imai, Tadayuki and Miyazu, Jun and Okabe, Yuichi and Ueno, Masahiro and Kobayashi, Junya},
    title = {Injected-charge-driven increase in electro-optic effect of KTN crystals},
    journal = {AIP Advances},
    volume = {4},
    number = {5},
    pages = {057109},
    year = {2014},
    month = {05},
    issn = {2158-3226},
    doi = {10.1063/1.4876237},
    url = {https://doi.org/10.1063/1.4876237},
    eprint = {https://pubs.aip.org/aip/adv/article-pdf/doi/10.1063/1.4876237/13134825/057109_1_online.pdf},
}

@misc{jayakumar2025,
      title={Relayed-loop optical scan amplification}, 
      author={Harishankar Jayakumar and Christopher Warkentin and Deano Farinella and Samuel Stanek and Zachary L Newman and Runze Liu and Aaron Kerlin},
      year={2025},
      eprint={2509.18399},
      archivePrefix={arXiv},
      primaryClass={physics.optics},
      url={https://arxiv.org/abs/2509.18399}, 
      doi = {10.48550/arXiv.2509.18399},
}

@article {yu2025,
	author = {Yu, Che-Hang and Yu, Yiyi and Canzano, Joseph S. and Fei, Yuandong and Smith, Spencer LaVere},
	title = {Non-inertial scan angle multiplier for expanded fields-of-view},
	elocation-id = {2025.06.13.659647},
	year = {2025},
	doi = {10.1101/2025.06.13.659647},
	publisher = {Cold Spring Harbor Laboratory},
	URL = {https://www.biorxiv.org/content/early/2025/06/15/2025.06.13.659647},
	eprint = {https://www.biorxiv.org/content/early/2025/06/15/2025.06.13.659647.full.pdf},
	journal = {bioRxiv}
}

@book{neamen2012,
  author    = {Donald A. Neamen},
  title     = {Semiconductor Physics and Devices: Basic Principles},
  edition   = {4},
  publisher = {McGraw--Hill},
  year      = {2012},
  address   = {New York}
}

@article{jaros1980,
author = {M. Jaros},
title = {Deep levels in semiconductors},
journal = {Advances in Physics},
volume = {29},
number = {3},
pages = {409--525},
year = {1980},
publisher = {Taylor \& Francis},
doi = {10.1080/00018738000101396},
URL = { 
    
        https://doi.org/10.1080/00018738000101396
},
}

@article{yagi2014,
author = {Yagi, Shogo and Fujiura, Kazuo},
year = {2014},
month = {12},
pages = {40–47},
title = {Electro-optic KTN devices},
volume = {56},
journal = {Physics Procedia},
doi = {10.1016/j.phpro.2014.08.093}
}

@article{herraez2002,
author = {Miguel Arevallilo Herr\'{a}ez and David R. Burton and Michael J. Lalor and Munther A. Gdeisat},
journal = {Appl. Opt.},
number = {35},
pages = {7437--7444},
publisher = {Optica Publishing Group},
title = {Fast two-dimensional phase-unwrapping algorithm based on sorting by reliability following a noncontinuous path},
volume = {41},
month = {Dec},
year = {2002},
url = {https://opg.optica.org/ao/abstract.cfm?URI=ao-41-35-7437},
doi = {10.1364/AO.41.007437},
}

@article{hariharan1987,
author = {P. Hariharan and B. F. Oreb and T. Eiju},
journal = {Appl. Opt.},
number = {13},
pages = {2504--2506},
publisher = {Optica Publishing Group},
title = {Digital phase-shifting interferometry: a simple error-compensating phase calculation algorithm},
volume = {26},
month = {Jul},
year = {1987},
url = {https://opg.optica.org/ao/abstract.cfm?URI=ao-26-13-2504},
doi = {10.1364/AO.26.002504},
}

@article{lu2019,
  author       = {Lu, D. and Zheng, Y. and Yuan, L.},
  title        = {Electron Paramagnetic Resonance Study on Oxygen Vacancies and Site Occupations in {Mg}-Doped {BaTiO3} Ceramics},
  journal      = {Materials},
  year         = {2019},
  volume       = {12},
  number       = {9},
  pages        = {1525},
  month        = may,
  doi          = {10.3390/ma12091525},
  pmid         = {31075960},
  pmcid        = {PMC6539086},
}

@article{dobaczewski2004,
    author = {Dobaczewski, L. and Peaker, A. R. and Bonde Nielsen, K.},
    title = {Laplace-transform deep-level spectroscopy: The technique and its applications to the study of point defects in semiconductors},
    journal = {Journal of Applied Physics},
    volume = {96},
    number = {9},
    pages = {4689-4728},
    year = {2004},
    month = {11},
    issn = {0021-8979},
    doi = {10.1063/1.1794897},
    url = {https://doi.org/10.1063/1.1794897},
    eprint = {https://pubs.aip.org/aip/jap/article-pdf/96/9/4689/18718882/4689_1_online.pdf},
}

@article{ivaylo2013,
author = {Ivaylo P. Ivanov and Xiangping Li and Philip R. Dolan and Min Gu},
journal = {Opt. Lett.},
number = {8},
pages = {1358--1360},
publisher = {Optica Publishing Group},
title = {Nonlinear absorption properties of the charge states of nitrogen-vacancy centers in nanodiamonds},
volume = {38},
month = {Apr},
year = {2013},
url = {https://opg.optica.org/ol/abstract.cfm?URI=ol-38-8-1358},
doi = {10.1364/OL.38.001358},
}

@INPROCEEDINGS{mendoza-lopez2022,
  author={Mendoza-Lopez, D. and Teyssedre, G. and Berquez, L. and Boudou, L.},
  booktitle={2022 IEEE 4th International Conference on Dielectrics (ICD)}, 
  title={Study of Trapping Process in BOPP by Coupled Space Charge and Photo-stimulated Discharge Techniques}, 
  year={2022},
  volume={},
  number={},
  pages={376-379},
  doi={10.1109/ICD53806.2022.9863568}}

@INPROCEEDINGS{zhu2009,
  author={Zhu, Zhien and Zhang, Yewen and An, Zhenlian and Zheng, Feihu},
  booktitle={2009 IEEE 9th International Conference on the Properties and Applications of Dielectric Materials}, 
  title={Study on charge trap distribution in LLDPE by photo-stimulated discharge current}, 
  year={2009},
  volume={},
  number={},
  pages={930-933},
  doi={10.1109/ICPADM.2009.5252249}}

@INPROCEEDINGS{boudou2018,
  author={Boudou, L. and Zheng, F. and Teyssedre, G.},
  booktitle={2018 12th International Conference on the Properties and Applications of Dielectric Materials (ICPADM)}, 
  title={Photo-stimulated discharge current measurements on biaxially oriented polypropylene thin films (BOPP)}, 
  year={2018},
  volume={},
  number={},
  pages={722-725},
  doi={10.1109/ICPADM.2018.8401125}}

@misc{stanek2024,
  author       = {Stanek, Samuel},
  title        = {Optical Detrapping in Potassium Tantalate Niobate Crystals},
  year         = {2024},
  howpublished = {University Digital Conservancy},
  url          = {https://hdl.handle.net/11299/270041},
  note         = {Retrieved from the University Digital Conservancy}
}

@ARTICLE{zheng2017,
  author={Zheng, Feihu and Li, Weimin and Gu, Min and An, Zhenlian and Zhang, Yewen},
  journal={IEEE Transactions on Dielectrics and Electrical Insulation}, 
  title={Photo-stimulated discharge current in polyimide films}, 
  year={2017},
  volume={24},
  number={3},
  pages={1802-1808},
  doi={10.1109/TDEI.2017.006372}}

@article{teyssedre_2021,
doi = {10.1088/1361-6463/abf44a},
url = {https://doi.org/10.1088/1361-6463/abf44a},
year = {2021},
month = {apr},
publisher = {IOP Publishing},
volume = {54},
number = {26},
pages = {263001},
author = {Teyssedre, Gilbert and Zheng, Feihu and Boudou, Laurent and Laurent, Christian},
title = {Charge trap spectroscopy in polymer dielectrics: a critical review},
journal = {Journal of Physics D: Applied Physics}
}

\end{document}